\documentclass[epj]{svjour}

\usepackage{graphicx}

\newcommand{\beq}{\begin{equation}}
\newcommand{\eeq}{\end{equation}}
\newcommand{\bfa}{\mbox{\boldmath $a$}}

\newcommand{\bfk}{\mbox{\boldmath $k$}}
\newcommand{\bfn}{\mbox{\boldmath $n$}}
\newcommand{\bfq}{\mbox{\boldmath $q$}}
\newcommand{\bfv}{\mbox{\boldmath $v$}}
\newcommand{\bfW}{\mbox{\boldmath $W$}}
\newcommand{\bfx}{\mbox{\boldmath $x$}}
\newcommand{\bfA}{\mbox{\boldmath $A$}}
\newcommand{\bfQ}{\mbox{\boldmath $Q$}}
\newcommand{\bfxi}{\mbox{\boldmath $\xi$}}
\newcommand{\ex}{\mbox{{\boldmath $\hat{e}$}}_{1}}
\newcommand{\ey}{\mbox{{\boldmath $\hat{e}$}}_{2}}
\newcommand{\ez}{\mbox{{\boldmath $\hat{e}$}}_{3}}
\newcommand{\bnabla}{\mbox{\boldmath $\nabla$}}
\newcommand{\cendot}{\mbox{\boldmath $\cdot$}}

\begin{document}

\title{Plane shearing waves of arbitrary form: exact solutions of 
the Navier--Stokes equations}
\author{Nishant K. Singh\inst{1,2,}\thanks{e-mail: singh@mps.mpg.de}
\and S. Sridhar\inst{3,}\thanks{e-mail: ssridhar@rri.res.in}
}%
\institute{$^{1}$Nordita, KTH Royal Institute of Technology and Stockholm
University, Roslagstullsbacken 23, SE-10691 Stockholm, Sweden\\
$^{2}$Max Planck Institute for Solar System Research,
Justus-von-Liebig-Weg 3, D-37077 G\"ottingen, Germany\\
$^{3}$Raman Research Institute, Sadashivanagar, Bangalore
560 080, India
}



\abstract{
We present exact solutions of the incompressible Navier--Stokes 
equations in a background linear shear flow. The method of 
construction is based on Kelvin's investigations into linearized 
disturbances in an unbounded Couette flow. We obtain explicit 
formulae for all three components of a Kelvin mode in terms of 
elementary functions. We then prove that Kelvin modes with 
parallel (though time--dependent) wave vectors can be superposed 
to construct the most general plane transverse shearing wave. 
An explicit solution is given, with any specified initial 
orientation, profile and polarization structure, with either 
unbounded or shear--periodic boundary conditions.
}

%

\maketitle
\section{Introduction}

In 1986 Craik and Criminale \cite{CC86} presented a class of exact 
solutions of the Navier--Stokes equations which were wavelike disturbances 
in background shear flows. Since then these solutions have proved 
extremely useful in the study of astrophysical and atmospheric fluid dynamics; 
a very useful collection of exact solutions can be found in \cite{DR06}. 
The approach taken in \cite{CC86} was a generalization of a century--old 
method invented by Kelvin \cite{Kel1887} to study linearized perturbations 
of Couette flows; see also \cite{MP77}. These shearing wave solutions, 
also referred to as \emph{Kelvin modes}, have time--dependent wave vectors 
and amplitudes. This feature makes them extremely useful in local 
stability analysis \cite{LH91,EY95}. Although a single Kelvin mode is 
an exact solution of the full Navier--Stokes (NS) equations, it has been 
remarked \cite{CC86} that until about 1965 there seems to be no evidence 
that this was so recognized; in fact, the first published mention is as 
late as 1983 \cite{Tun83}. Moreover, an explicit formula has been 
published \cite{Kel1887,CC86} for only one of the three components of 
the disturbance. 

In this paper we present exact solutions for all three components of 
the velocity field of a Kelvin mode, in closed form using only elementary 
mathematical functions. We identify a subset of these modes whose wave 
vectors --- though time--dependent --- remain parallel to each other for 
all time. These are used to synthesize the most general plane transverse 
shearing wave, which can have any specified initial orientation, profile 
and polarization structure, with either unbounded or shear--periodic 
boundary conditions.

Let $(\ex,\ey,\ez)$ be the unit basis vectors of a Cartesian coordinate 
system in the laboratory frame. Using notation $\bfx = (x_1,x_2,x_3)$ 
for the position vector and $t$ for time, we write the total fluid 
velocity as $(Sx_1\ey + \bfv)$, where $S$ is the rate of shear 
parameter and $\bfv(\bfx, t)$ is the incompressible disturbance 
($\bnabla\cendot\bfv = 0$) which obeys the NS equations:

\begin{eqnarray}
\left(\partial_t \;+\; Sx_1\partial_2\right)\bfv  &\;+\;& Sv_1\ey 
\;+\; \left(\bfv\cendot\bnabla\right)\bfv \;=\; 
- \bnabla p \;+\; \nu \nabla^2 \bfv\,,\nonumber\\[1em]
\nabla^2 p &\;=\;& - \bnabla\cendot\left[\left(\bfv\cendot\bnabla\right)
\bfv \right] \;-\; 2S\partial_2 v_1\,. 
\label{NSeq}
\end{eqnarray}

\noindent
We seek a solution in the form of a single Kelvin mode 

\begin{eqnarray}
\bfv_{\bfk}(\bfx, t) \;&=&\; {\rm Re}\left\{\bfA(\bfk, t)\,
\exp{\left[{\rm i}\,\bfk^{\rm sh}(t)\cendot \bfx \right]}
\right\}\,, \nonumber \\[2ex]
p_{\bfk}(\bfx, t) \;&=&\; {\rm Re}\left\{\psi(\bfk, t)\, 
\exp{\left[ {\rm i}\,\bfk^{\rm sh}(t)\cendot \bfx \right]} \right\}\,,
\label{singplwave}
\end{eqnarray}

\noindent
where the time--dependent sheared wave vector, $\bfk^{\rm sh}(t)$, has components

\beq
k^{\rm sh}_1 \;=\; k_1 - St k_2\,, \qquad k^{\rm sh}_2 \;=\; k_2\,, \qquad 
k^{\rm sh}_3 \;=\; k_3\,,
\label{inshtr}
\eeq

\noindent 
with $\bfk \equiv (k_1, k_2, k_3)$ being a constant wave vector. Our task 
now is to determine the amplitudes $\bfA(\bfk, t)$. Incompressibility 
requires that $\bfk^{\rm sh}(t)\cendot\bfA(\bfk,t) \,=\, 0\,$. Therefore, 
when eqns.~(\ref{singplwave}) and (\ref{inshtr}) are substituted in 
eqns.~(\ref{NSeq}), the nonlinear term, $\left(\bfv\cendot\bnabla\right)\bfv$ 
vanishes because
  
\beq
\left(\bfA\cendot\bnabla\right)\, \exp{\left[{\rm i}\,\bfk^{\rm sh}(t)
\cendot \bfx \right]} \;=\; \left({\rm i}\,\bfk^{\rm sh}(t)\cendot\bfA
\right)\,\exp{\left[{\rm i}\,\bfk^{\rm sh}(t)\cendot \bfx \right]} \;=\; 0\,.
\label{nonlinvanish}
\eeq

\noindent
The pressure can be eliminated by using the second of eqns.~(\ref{NSeq}):    
$\left|\bfk^{\rm sh}(t)\right|^2 \psi \;=\; 2{\rm i}Sk_2A_1\,$, where 
$\left|\bfk^{\rm sh}(t)\right|^2 \;=\; 
\left[\left(k_1 - St k_2\right)^2 + k_2^2 + k_3^2\right]$. 
Then $\bfA$ satisfies

\beq
\partial_t\bfA \;+\; SA_1 \ey \;=\; 
2S\left(\frac{k_2 \bfk^{\rm sh}(t)}{\left|\bfk^{\rm sh}(t)\right|^2} \right)A_1 
\;-\; \nu\left|\bfk^{\rm sh}(t)\right|^2 \bfA\,.
\label{Aeq}
\eeq

\noindent
We now obtain explicit solutions for $\bfA$. To do this, define a 
new amplitude variable, $\bfa(\bfk, t)$, by

\beq
\bfA(\bfk, t) \;=\; \widetilde{G}_{\nu}(\bfk, t)\,\bfa(\bfk, t)\,,
\label{A-a}
\eeq

\noindent 
where $\widetilde{G}_{\nu}(\bfk, t)$ is a Fourier--space viscous Green's function,

\beq
\widetilde{G}_{\nu}(\bfk, t) \;=\; \exp{\left[-\nu \int_{0}^{t} \, 
\mathrm{d}s \, \left|\bfk^{\rm sh}(s)\right|^2 \right]} \;=\; 
\exp{\left[-\nu\left(k^2t \,-\, Sk_1k_2t^2 \,+\, 
\frac{S^2}{3}k_2^2t^3\right)\right]}\,.
\label{greenfn}
\eeq

\noindent
When eqns.~(\ref{A-a}) and (\ref{greenfn}) substituted in eqn.~(\ref{Aeq}), 
we obtain the following equations for the three components of $\bfa(\bfk, t)\,$:

\begin{eqnarray}
\label{a1}
\partial_t a_1 \;-\; 2S\left[\frac{\left(k_1 - Stk_2\right) k_2}
{\left(k_1 - St k_2\right)^2 + k_2^2 + k_3^2} \right]\,a_1 \;&=&\; 0\,, \\[2ex]
\label{a2}
\partial_t a_2 \;-\; 2S\left[\frac{k_2^2}
{\left(k_1 - St k_2\right)^2 + k_2^2 + k_3^2} \,-\, \frac{1}{2} \right]
\,a_1 \;&=&\; 0\,, \\[2ex]
\label{a3}
\partial_t a_3 \;-\; 2S\left[\frac{k_2 k_3}{\left(k_1 - St k_2\right)^2 
+ k_2^2 + k_3^2} \right]\,a_1 \;&=&\; 0\,.
\end{eqnarray}

\noindent
Eqn.~(\ref{a1}) can be solved to get an explicit expression for $a_1(\bfk, t)$:

\beq
a_1(\bfk, t) \;=\; \frac{k^2}{\left(k_1 - St k_2\right)^2 + k_2^2 + k_3^2}\,
a_1(\bfk, 0)\,,
\label{solna1}
\eeq 

\noindent
which is given in \cite{Kel1887}. When this is substituted 
in eqns.~(\ref{a2}) and (\ref{a3}), the latter can be integrated to 
obtain expressions for $a_2(\bfk, t)$ and $a_3(\bfk, t)$. However, 
neither Kelvin nor anyone else, to the best of our knowledge, have 
published explicit formulae for these two 
components.\footnote{Markus and Press \cite{MP77}
study perturbations of plane Couette 
flow using Kelvin waves. However, their analysis is limited to two 
dimensional perturbations, whereas the shearing waves we consider here are 
fully three dimensional.} Thus we were pleasantly surprised to 
find that $a_2(\bfk, t)$ and $a_3(\bfk, t)$ could be expressed entirely 
in terms of elementary functions: 

\begin{eqnarray}
a_2(\bfk, t) &\;=\;& a_2(\bfk, 0) \;+\; \left\{\frac{k^2k_3^2}{k_2\left(k_2^2 
+ k_3^2\right)^{3/2}}\left[\arctan\left(\frac{k_1 - Stk_2}
{\sqrt{k_2^2 + k_3^2}}\right) \,-\, \arctan\left(\frac{k_1}
{\sqrt{k_2^2 + k_3^2}}\right)\right]\right.\nonumber\\[2ex]
&&\left.\qquad\qquad \;-\;\;\; \frac{k^2k_2}{k_2^2 + k_3^2}
\left[\frac{k_1 - Stk_2}{\left(k_1 - St k_2\right)^2 + k_2^2 + k_3^2} 
\;-\; \frac{k_1}{k^2}\right]\right\}\,a_1(\bfk, 0)\,,
\label{solna2}\\[2em]
a_3(\bfk, t) &\;=\;& a_3(\bfk, 0) \;-\; \left\{\frac{k^2k_3}
{\left(k_2^2 + k_3^2\right)^{3/2}}\left[\arctan\left(\frac{k_1 - Stk_2}
{\sqrt{k_2^2 + k_3^2}}\right) \,-\, \arctan
\left(\frac{k_1}{\sqrt{k_2^2 + k_3^2}}\right)\right]\right.\nonumber\\[2ex]
&&\left.\qquad\qquad \;+\;\;\; \frac{k^2k_3}{k_2^2 + k_3^2}
\left[\frac{k_1 - Stk_2}{\left(k_1 - St k_2\right)^2 + k_2^2 + k_3^2} 
\,-\, \frac{k_1}{k^2}\right]\right\}\,a_1(\bfk, 0)\,.
\label{solna3}
\end{eqnarray}

\begin{figure}
\begin{center}
\includegraphics[scale=0.7]{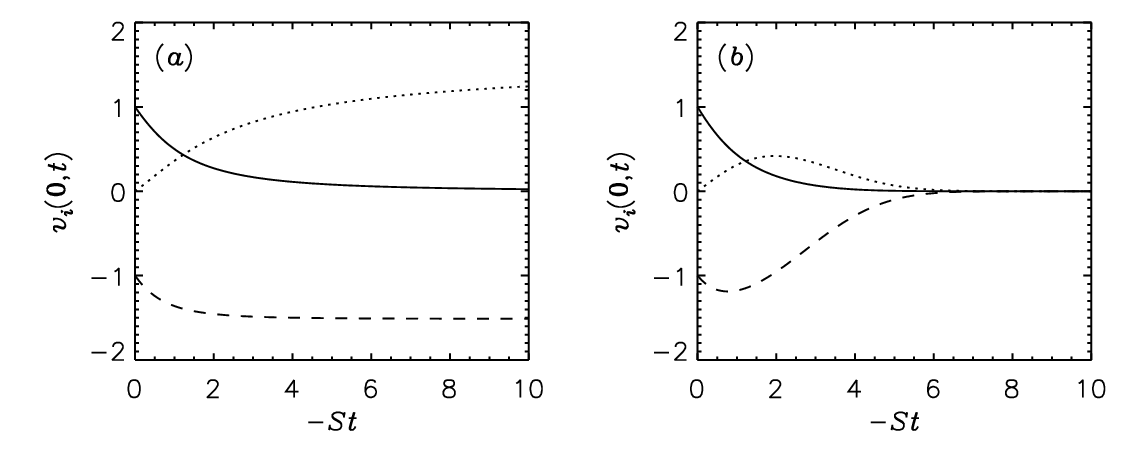}
\caption{Plots of the three components of the velocity field, 
measured at the origin, as functions of $St$. We have chosen 
$\bfk = (1,1,1)$ and $\bfa(\bfk, 0) = (1, 0, -1)$. The bold lines 
are for $v_1({\bf 0}, t)$, the dotted for 
$v_2({\bf 0}, t)$, and the dashed for $v_3({\bf 0}, t)$. 
Panel~(a) is for the non viscous case, $\nu = 0$, so the velocity 
components are identical to the amplitudes, $\bfa(\bfk, t)$, 
of eqns.~(\ref{solna1})---(\ref{solna3}). Panel~(b) corresponds 
to  $(\nu k^2/S) = -0.1$, and all three components ultimately 
suffer viscous decay.}
\label{velcomp}
\end{center}
\end{figure}

\noindent
Incompressibility requires that $\bfk^{\rm sh}(t)\cendot\bfa(\bfk, t) \,=\, 0$, 
which is guaranteed if the initial conditions are chosen such that 
$\bfk\cendot\bfa(\bfk, 0) \,=\, 0\,$. From eqns.~(\ref{solna1})---(\ref{solna3}), 
we can see that, at late times, $a_1(\bfk, t)\to 0$, whereas both 
$a_2(\bfk, t)$ and $a_3(\bfk, t)$ saturate at non zero values. 
This happens because the background flow shears out the $a_1$ 
component, and generates the $a_2$ and $a_3$ components.

\begin{figure}
\begin{center}
\includegraphics[scale=0.23]{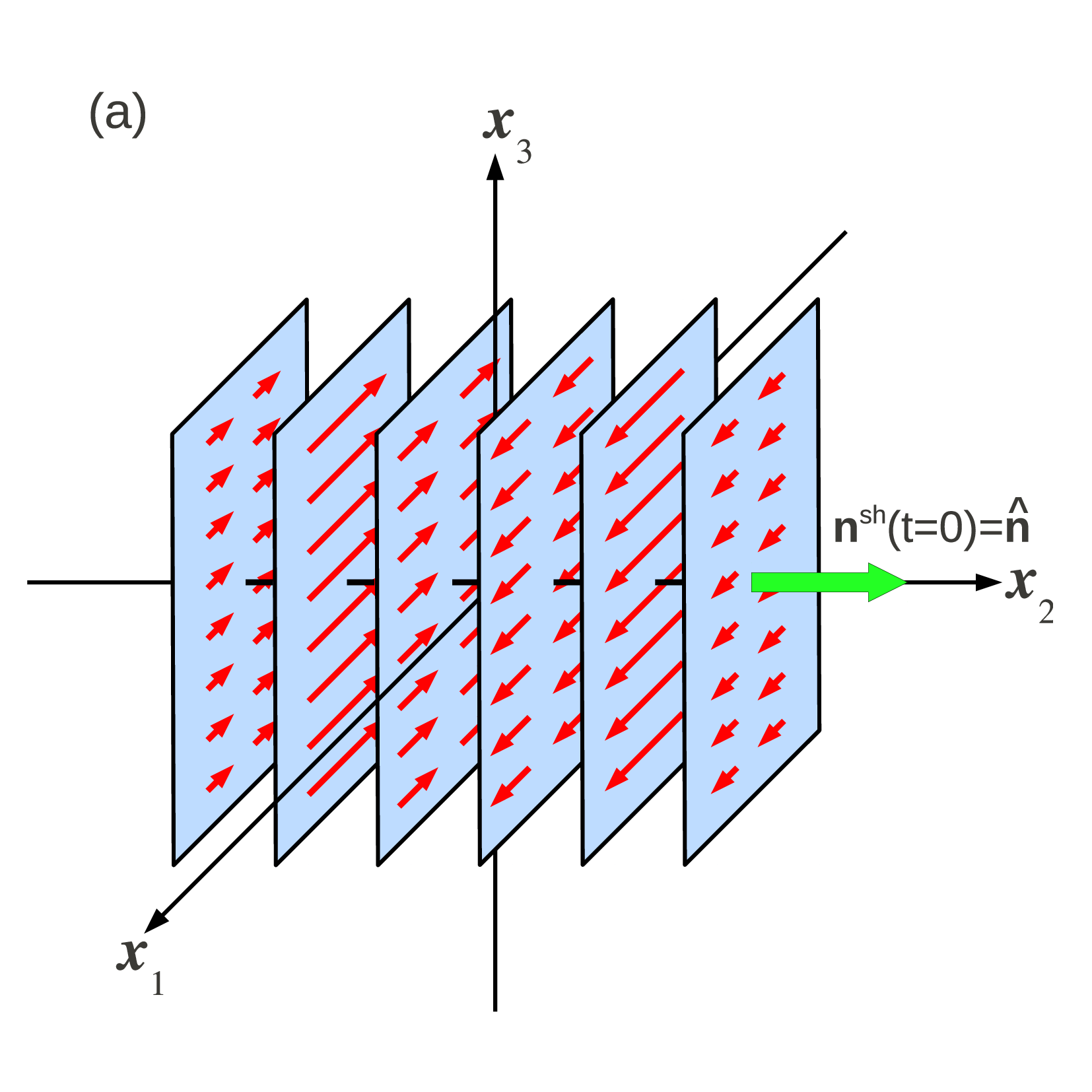}
\hskip0.1cm
\includegraphics[scale=0.23]{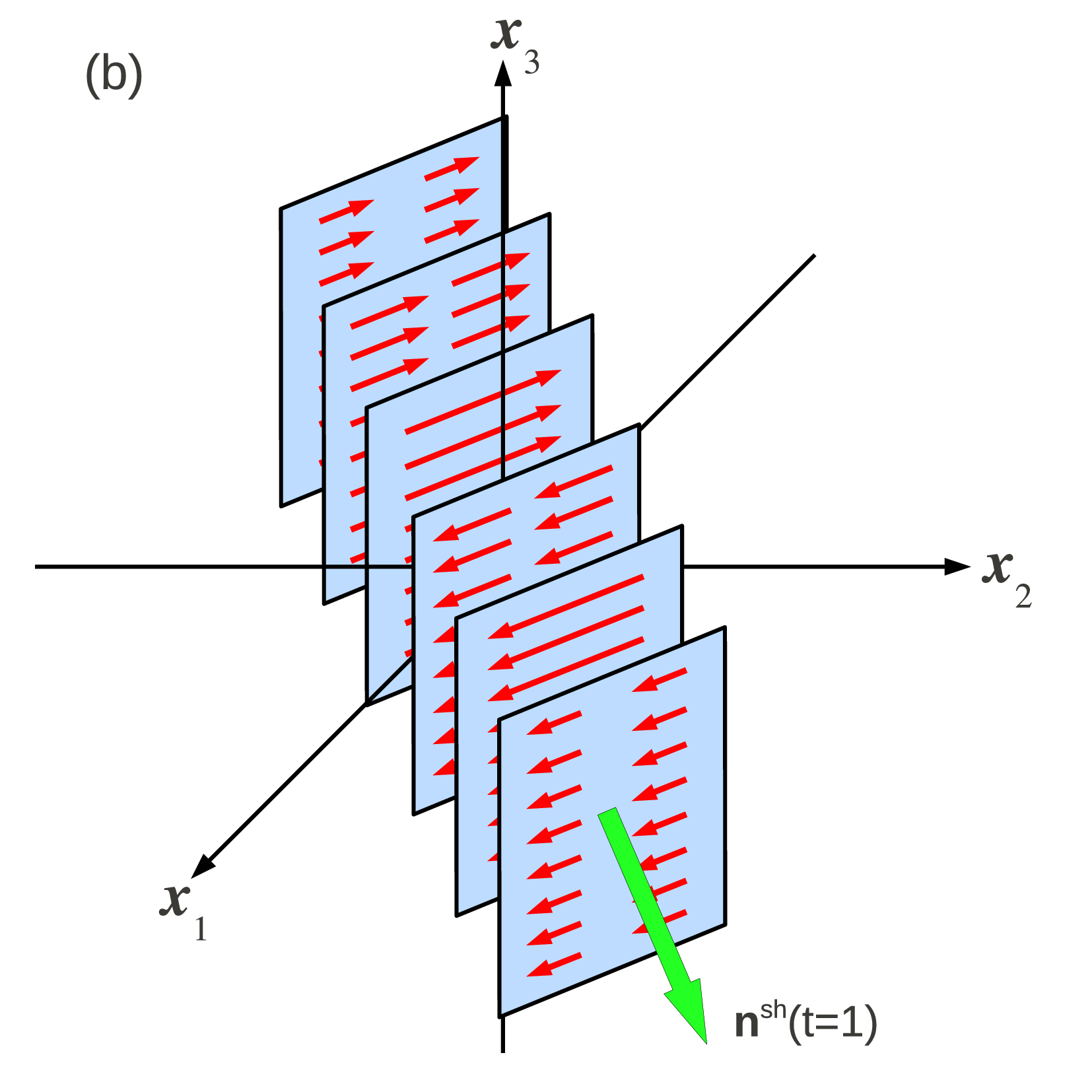}
\vskip1.0cm 
\includegraphics[scale=0.23]{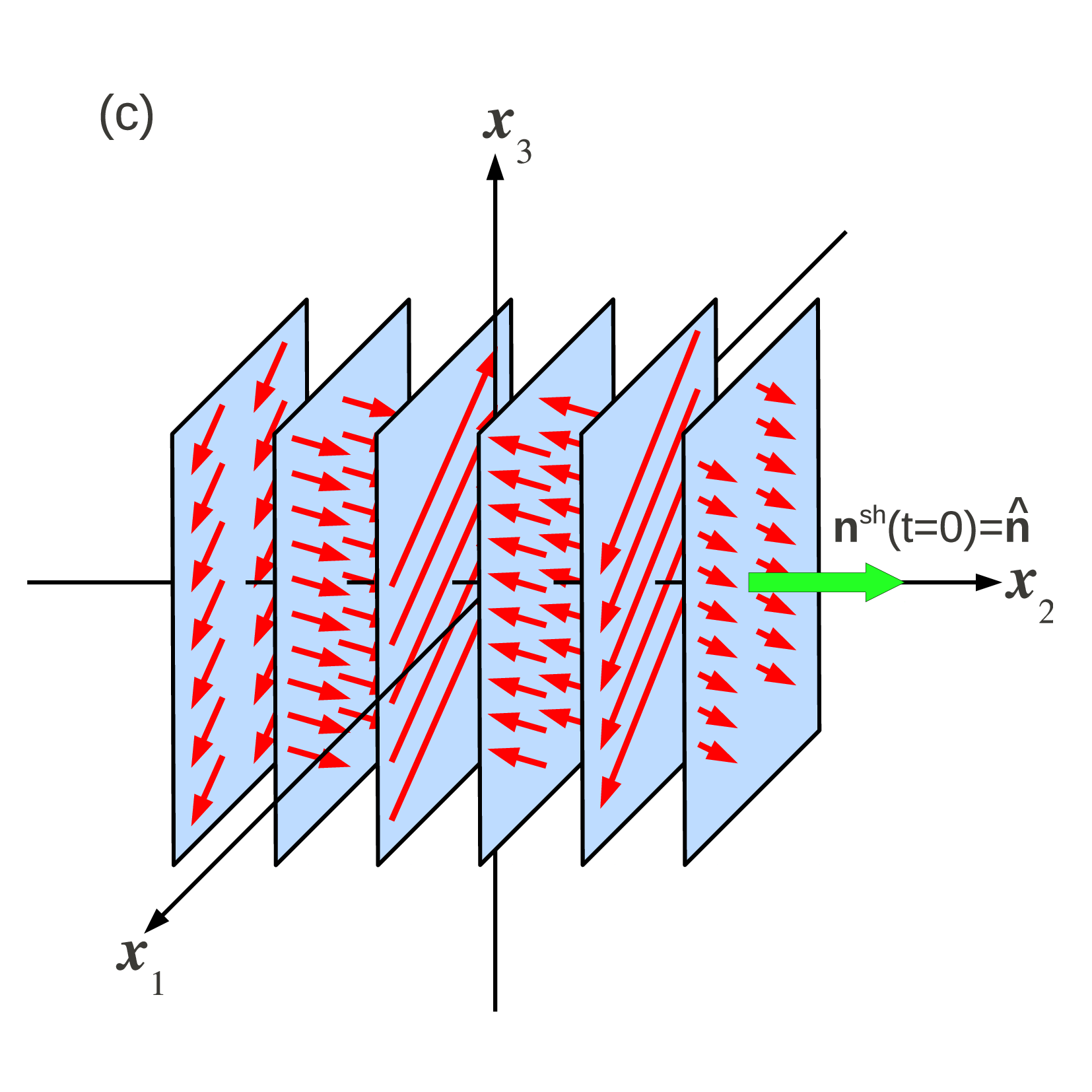}
\hskip0.1cm
\includegraphics[scale=0.23]{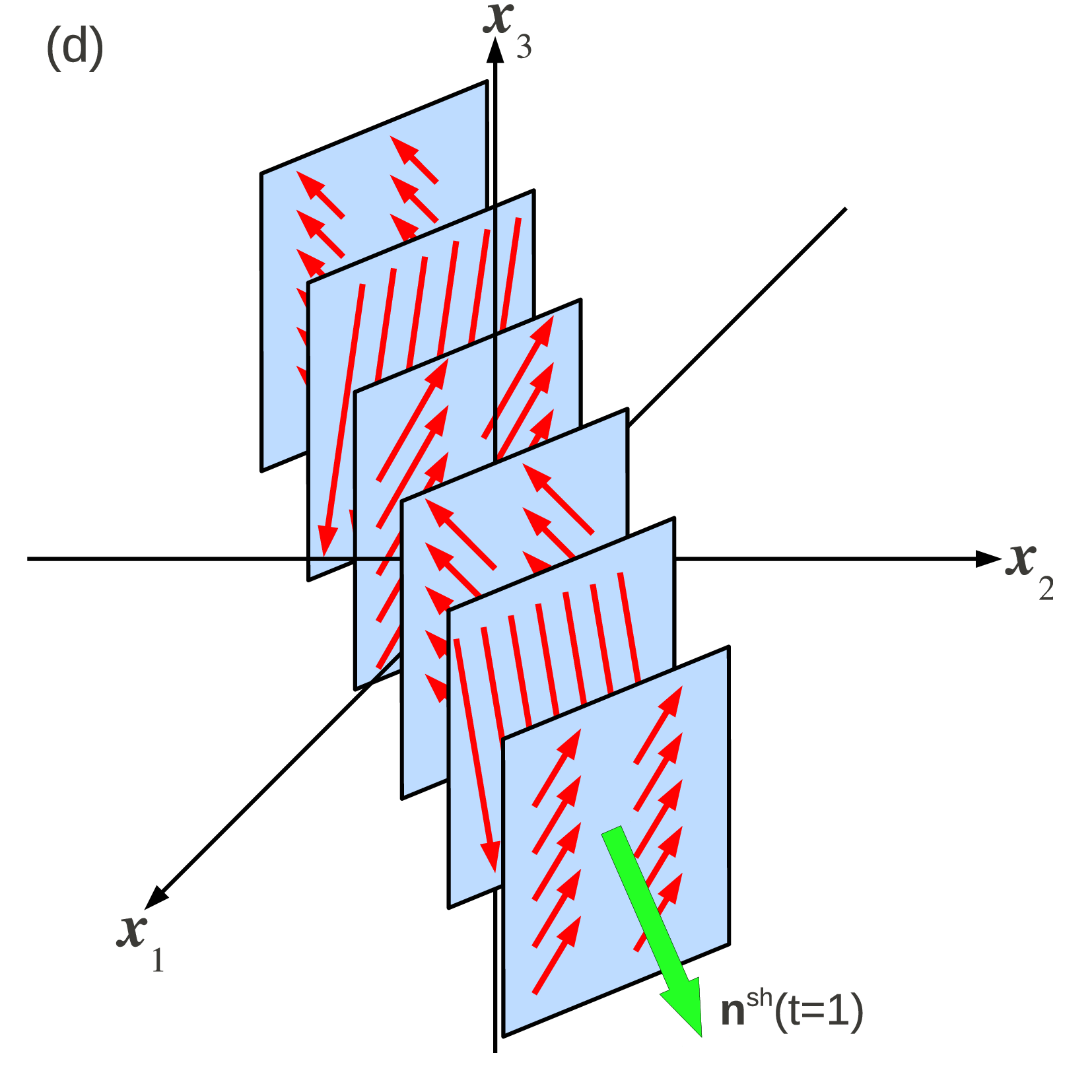}
\caption{(color online) Evolution of plane transverse shearing wavepackets.
The polarization structure of the velocity field is indicated on some 
sections of the plane wavefronts. The parameters values used are 
$S=-1$, $\nu = 1$, $W_0=1$, $\sigma=10$ and $k=1$. (a) and (b) 
Linearly polarized ($h=0$) wavepackets at times $t=0$ and $t=1$. 
(c) and (d) Right circularly polarized ($h=1$) at times $t=0$ and $t=1$.}
\label{shearwaves}
\end{center}
\end{figure}

When eqns.~(\ref{A-a}), (\ref{greenfn}), (\ref{solna1})---(\ref{solna3}) 
are substituted in eqn.~(\ref{singplwave}), we obtain the full velocity 
field of a single Kelvin mode; it is readily verified that structure 
of the mode depends on the dimensionless variable, $St$, and the 
dimensionless parameter, $(\nu k^2/S)$. The spatio--temporal behavior 
of these modes is briefly explored through Figs.~(\ref{velcomp}) and 
(\ref{shearwaves}). In order to understand its time variation, it is 
convenient to measure the velocity components at the origin, as is done 
in Fig.~(\ref{velcomp}). Then, 
$\bfv({\bf 0}, t) = \widetilde{G}_{\nu}(\bfk, t)\,{\rm Re}\{\bfa(\bfk, t)\}$. 
Fig.~(1a) corresponds to the case of zero viscosity, $(\nu k^2/S) = 0$. 
In this case $\widetilde{G}_{\nu}=1$, and the plots give 
$\bfv({\bf 0}, t) = {\rm Re}\{\bfa(\bfk, t)\}$, where we can see the 
decay of $a_1$ and the saturation of $a_2$ and $a_3$ discussed above. 
In Fig.~(1b), we have chosen $(\nu k^2/S) = -0.1$, so that all three 
components of $\bfv({\bf 0}, t)$ ultimately suffer viscous decay. 
It can be seen that, before this decay, there is transient amplification 
of $v_2$ and $v_3$, due to competition between shear and viscosity. For 
larger values of viscosity (not shown here), this transient amplification 
may be absent because the damping can overwhelm shear.     

Until now we have considered an unbounded flow. However, in numerical 
simulations of the local dynamics of differentially rotating discs in 
astrophysical systems \cite{BT08,BH98}, it is customary to 
employ ``shear--periodic'' boundary conditions. Let us define 
\emph{sheared coordinates} by 

\beq
x^{\rm sh}_1 = x_1\,,\qquad x^{\rm sh}_2 = x_2 - St x_1\,,\qquad x^{\rm sh}_3 = x_3\,.
\label{sheartr}
\eeq

\noindent
These may be thought of as the Lagrangian coordinates of fluid elements that 
are carried along by the background shear flow. A function is said 
to be \emph{shear--periodic} when it is a periodic function of 
$(x^{\rm sh}_1, x^{\rm sh}_2, x^{\rm sh}_3)$ with periodicities 
$(L_1, L_2, L_3)$, respectively. The phase of the function $\bfv_{\bfk}$ 
can be written as $\bfk^{\rm sh}(t)\cendot \bfx \,=\, \bfk\cendot\bfx^{\rm sh}$. 
Therefore, a shear--periodic Kelvin mode has wave vectors 
$\bfk\in\left(2\pi m_1/L_1, 2\pi m_2/L_2, 2\pi m_3/L_3\right)$, where the 
$m_i$ take any integer values. 

We now use the explicit expressions obtained for the Kelvin modes to
construct the most general plane transverse shearing wave. Let us 
consider two Kelvin modes, $\bfv_{\bfk}(\bfx, t)$ and 
$\bfv_{\bfk'}(\bfx, t)$ corresponding to wave vectors $\bfk$ and $\bfk'$, 
which are parallel to each other but could differ in magnitudes. 
Using eqns.~(\ref{inshtr}), we see that the corresponding sheared 
wave vectors, $\bfk ^{\rm sh}(t)$ and $\bfk^{'\rm sh}(t),$ are also 
parallel to each other for all time. Incompressibility implies that 
$\bfv_{\bfk}(\bfx, t)$ and $\bfv_{\bfk'}(\bfx, t)$ are perpendicular 
to $\bfk ^{\rm sh}(t)$ and $\bfk^{'\rm sh}(t)$ for all time. So, if we 
superpose $\bfv_{\bfk}(\bfx, t)$ and $\bfv_{\bfk'}(\bfx, t)$, the nonlinear 
term in the NS equations vanishes, because the superposed velocity field 
remains parallel to the wavefronts. Thus the superposition of an 
arbitrary number of Kelvin modes, all with wave vectors parallel to 
each other, is an exact solution of the NS equations. 

Let us choose a unit vector ${\bf \hat{\bfn}} \,=\, (n_1, n_2, n_3)$, and 
define the sheared (non--unit) vector $\bfn^{\rm sh}(t)$ by

\beq
n^{\rm sh}_1 \;=\; \left(n_1 - St n_2\right)\,, \qquad n^{\rm sh}_2 
\;=\; n_2\,, \qquad n^{\rm sh}_3 \;=\; n_3\,.
\label{qsh}
\eeq

\noindent
Superposing all Kelvin modes with wave vectors $\bfq \,=\, q{\bf \hat{\bfn}}$, 
where $-\infty < q < \infty$, we obtain an exact plane--wave solution of 
the NS equations with wavefronts perpendicular to $\bfn^{\rm sh}(t)$: 

\begin{eqnarray}
v_i(\bfx, t) &\;=\;& \int_{-\infty}^{\infty}\frac{{\rm d}q}{2\pi}\;
\widetilde{G}_{\nu}(q{\bf \hat{\bfn}}, t)\,\widetilde{W}_i(q)\,
\exp{\left[{\rm i}q\bfn^{\rm sh}(t)\cendot\bfx\right]} \;+\;\nonumber\\[2ex] 
&&\;+\; \left[\frac{F_i\left(\bfn^{\rm sh}(t)\right) - 
F_i({\bf \hat{\bfn}})}{n_2^2 + n_3^2}\right]
\int_{-\infty}^{\infty}\frac{{\rm d}q}{2\pi}\;
\widetilde{G}_{\nu}(q{\bf \hat{\bfn}}, t)\,\widetilde{W}_1(q)\,
\exp{\left[{\rm i}q\bfn^{\rm sh}(t)\cendot\bfx\right]},
\label{plane}
\end{eqnarray}

\noindent
where the dimensionless and scale--invariant functions, $F_i(\bfQ)$, are defined by

\beq
F_i(\bfQ) \;=\; \frac{Q_3}{\sqrt{Q_2^2 + Q_3^2}}
\left[\frac{Q_3}{Q_2}\,\delta_{i2} \;-\; \delta_{i3}\right] 
\arctan{\left(\frac{Q_1}{\sqrt{Q_2^2 + Q_3^2}}\right)} \;-\; \frac{Q_1Q_i}{Q^2}\,. 
\label{Fdef}
\eeq

\noindent
For shear--periodic boundary conditions, the integral 
over $q$ in eqn.~(\ref{plane}) should be replaced by an appropriate sum. 
The $\widetilde{\bfW}(q)$ are Fourier--space initial conditions 
corresponding to the $\bfa(\bfk, 0)$ of eqns.~(\ref{solna1})---(\ref{solna3}), 
and must satisfy the incompressibility condition, 
${\bf \hat{\bfn}}\cendot\widetilde{\bfW}(q) \,=\, 0\,$. They are 
determined by the initial profile and polarization stucture of the 
plane wave. At $t=0$, the wavefronts are perpendicular 
to ${\bf \hat{\bfn}}$, so we write 
$\bfv(\bfx, 0) \,=\, \bfW({\bf \hat{\bfn}}\cendot\bfx)$, where 
${\bf \hat{\bfn}}\cendot\bfW \,=\, 0$. Note that the only constraint 
on the initial condition, $\bfW$, is that it is a vector field 
that is perpendicular everywhere to the unit vector ${\bf \hat{\bfn}}$; 
otherwise it is a quite arbitrary function of its one argument. Thus, 
no restriction need be placed on the initial profile and polarization 
structure of the initial conditions. Given $\bfW(y)$, we can 
determine $\widetilde{\bfW}(q) = \int_{-\infty}^{\infty}{\rm d}y\,
\bfW(y)\exp{\left[-{\rm i}qy\right]}\,$, and use this in eqn.~(\ref{plane}) 
to calculate $\bfv(\bfx, t)$. 

Eqn.~(\ref{plane}) is a complete solution for a general plane shearing wave, 
expressed in terms of a Fourier integral. However, it is physically 
more transparent to rewrite the right side in terms of real--space quantities. 
To do this, we must introduce the real--space viscous Green's function, 
whose natural definition is with respect to the sheared coordinates \cite{SS10}:

\beq
G_\nu\left(\bfx^{\rm sh}, t\right) \;=\; \int\frac{{\rm d}\bfk}{(2\pi)^3}\,
\widetilde{G}_{\nu}(\bfk, t)\,\exp{\left[{\rm i}\bfk\cendot\bfx^{\rm sh}\right]}\,.
\label{greenfnre}
\eeq
 
\noindent
The properties of this function are discussed in \cite{KR71,SS10}, 
where it is shown that it takes the form of a sheared heat kernel, which 
is an anisotropic Gaussian function of $\bfx^{\rm sh}$ with time--dependent 
coefficients; all the principal axes increase without bound and rotate 
against the direction of the background shear. Noting that 
$\bfn^{\rm sh}(t)\cendot \bfx \,=\, {\bf \hat{\bfn}}\cendot\bfx^{\rm sh}$, 
we can write the general form of the plane shearing wave as  

\begin{eqnarray}
v_i(\bfx, t) &\;=\;& \int{\rm d}^3\xi\,G_\nu(\bfxi, t)\,
W_i\left({\bf \hat{\bfn}}\cendot[\bfx^{\rm sh}(t) - \bfxi]\right) 
\;\;+\;\nonumber\\[2em]  
&&\;+\;\; \left[\frac{F_i\left(\bfn^{\rm sh}(t)\right) - 
F_i({\bf \hat{\bfn}})}{n_2^2 + n_3^2}\right]
\int{\rm d}^3\xi\,G_\nu(\bfxi, t)\,W_1\left({\bf \hat{\bfn}}\cendot
[\bfx^{\rm sh}(t) - \bfxi]\right)
\label{planers}
\end{eqnarray}

\noindent
As an illustrative example let us consider the following initial condition, 
corresponding to a polarized wavepacket with wave vector pointing along 
the $x_2$--axis: ${\bf \hat{\bfn}} = \ey$, $W_1(x_2) = 
W_0\exp{\left[-x_2^2/2\sigma^2\right]}\sin{kx_2}\,$, $W_2 =0\,$, 
$W_3(x_2) = hW_0\exp{\left[-x_2^2/2\sigma^2\right]}\cos{kx_2}\,$, 
where $-1\leq h\leq 1\,$. The wavepacket is linearly polarized when 
$h=0$, and right/left circularly polarized when $h = \pm 1$; other 
values of $h$ correspond to different degrees of elliptical polarizations. 
At a later time, the wave vector has components 
$n^{\rm sh}_1 = -St\,$, $n^{\rm sh}_2 = 1\,$, $n^{\rm sh}_3 =0\,$. 
Since both $W_i$ and $G_\nu(\bfxi, t)$ are Gaussian functions, the 
integrals in eqn.~(\ref{planers}) can be performed analytically and 
$\bfv(\bfx, t)$ evaluated explicitly. We present the results graphically 
in Fig.(\ref{shearwaves}) for two cases, one linearly polarized and 
the other right circularly polarized. As the wavepackets are sheared, 
they undergo transient amplification due to the combined action of 
shear and viscosity, and at late times suffer viscous damping. 

In conclusion, we have constructed exact solutions of the Navier--Stokes 
equation with a background linear shear flow. All three components of 
the velocity field of the Kelvin modes are given in closed form using 
only elementary mathematical functions. It is demonstrated that, when 
Kelvin modes with parallel wave vectors are superposed, they remain 
exact solutions. We give in explicit form the most general plane transverse 
shearing waves, with any specified initial orientation, profile and 
polarization structure, with either unbounded or shear--periodic 
boundary conditions.
Of particular interest is the stability of our solutions;
if they are stable then they might serve as local representations
of disturbances in simulations of astrophysical flows.

\begin{thebibliography}{99}

\bibitem{CC86}
A.~D.~D.~Craik \& W.~O.~Criminale, Proc. R. Soc. Lond. A \textbf{406}, 13
(1986).

\bibitem{DR06}
P.~G.~Drazin \& N.~Riley,
\textit{The Navier--Stokes equations: a classification of flows 
and exact solutions}. 
London Mathematical Society Lecture Note Series. 334, 
Cambridge University Press (2006).

\bibitem{Kel1887}
W.~Thomson (Lord Kelvin), Philosophical Magazine \textbf{24} (5), 188 (1887).

\bibitem{MP77}
P.~S.~Marcus \& W.~H.~Press, Journal of Fluid Mechanics \textbf{79},
525 (1977).

\bibitem{LH91}
A.~Lifschitz \& E.~Hameiri, Physics of Fluids A \textbf{3}, 2644 (1991).

\bibitem{EY95}
B.~Eckhardt \& D.~Yao, Chaos, Solitons and Fractals \textbf{5}, 2073 (1995).

\bibitem{Tun83}
K.~K.~Tung, Journal of Fluid Mechanics \textbf{133}, 443 (1983).

\bibitem{BT08}
J.~Binney \& S.~Tremaine, \textit{Galactic Dynamics: Second Edition}.
Princeton University Press (2008).  

\bibitem{BH98}
S.~A.~Balbus \& J.~F.~Hawley, Reviews of Modern Physics \textbf{70}, 1 (1998). 
 
\bibitem{SS10} 
S.~Sridhar \& N.~K.~Singh, Journal of Fluid Mechanics \textbf{664}, 265 (2010).

\bibitem{KR71} 
F.~Krause \& K.-H.~R\"adler,
Elektrodynamik der mittleren Felder in turbulenten 
leitenden Medien und Dynamotheorie. In \emph{Ergebnisse der 
Plasmaphysik und der Gaselektronik}, Band 2, pp.~2--154. Akademie (1971).
\end{thebibliography}
\end{document}